# Climatic seasonality may affect ecological network structure: Food webs and mutualistic networks

Kazuhiro Takemoto[1], Saori Kanamaru, Wenfeng Feng
*Department of Bioscience and Bioinformatics, Kyushu Institute of Technology, Kawazu 680-4, Iizuka, Fukuoka 820-8502, Japan*

**Abstract**
Ecological networks exhibit non-random structural patterns, such as modularity and nestedness, which indicate ecosystem stability, species diversity, and connectance. Such structure-stability relationships are well known. However, another important perspective is less well understood: the relationship between the environment and structure. Inspired by theoretical studies that suggest that network structure can change due to environmental variability, we collected data on a number of empirical food webs and mutualistic networks and evaluated the effect of climatic seasonality on ecological network structure. As expected, we found that climatic seasonality affects ecological network structure. In particular, an increase in modularity due to climatic seasonality was observed in food webs; however, it is debatable whether this occurs in mutualistic networks. Interestingly, the type of climatic seasonality that affects network structure differs with ecosystem type. Rainfall and temperature seasonality influence freshwater food webs and mutualistic networks, respectively; food webs are smaller, and more modular, with increasing rainfall seasonality. Mutualistic networks exhibit a higher diversity (particularly of animals) with increasing temperature seasonality. These results confirm the theoretical prediction that stability increases with greater perturbation. Although these results are still debatable because of several limitations in the data analysis, they may enhance our understanding of environment-structure relationships.

## 1. Introduction

Ecological communities consist of a number of species that are connected via interspecific interactions, such as trophic and mutualistic relationships. Their structure and dynamics are significant in ecology, because they are important not only in the context of basic scientific research, such as structure-stability relationships (Allesina and Tang, 2012; Bascompte, 2010; Mougi and Kondoh, 2012; Thébault and Fontaine, 2010), as typified by May's paradox (May, 1972), but also in the context of applied ecology, such as biodiversity maintenance and environmental sustainability (Allesina and Tang, 2012; Bascompte, 2010; Mougi and Kondoh, 2012).

---

The development of field observation technology, and the improvement of infrastructures such as databases, have increased the availability of ecological data on interspecific interactions and have enabled large-scale data analysis of real-world ecosystems. Because of the importance of network science (Barabási, 2013), ecological communities are often represented as networks (Bascompte, 2010; Proulx et al., 2005) (so called *ecological networks*, in which nodes and edges correspond to species and interspecific interactions, respectively), and have been actively investigated recently using complex network analysis techniques (Takemoto and Oosawa, 2012).

Previous network analytical studies have revealed that ecological networks (plant–animal mutualistic networks in particular) exhibit two representative non-random structural patterns: a modular structure (Olesen et al., 2007), and nested architecture (Bascompte et al., 2003). Despite the correlation between them, these two structural patterns can provide complementary information on how interactions are organised in communities (Fortuna et al., 2010). The modular or compartmentalised structure describes the deconstruction of a network into dense, and yet, weakly interconnected subnetworks (subgroups), and indeed, modular organisation is an important feature of biological systems (Hartwell et al., 1999). A nested structure indicates that the interaction pairs of a certain (specialist) species form a subset of those of another (generalist) species, in a hierarchical fashion. However, the degree of modular and nested architectures (generally called *modularity* and *nestedness*, respectively) differ with the ecosystem type: antagonistic (or trophic) networks (e.g. food webs) and mutualistic networks (e.g. plant-pollinator networks). In general, the modularity of antagonistic networks is higher than that of mutualistic networks, and the nestedness of antagonistic (i.e. resource-consumer) networks is lower than that of mutualistic networks (Bascompte et al., 2003; Thébault and Fontaine, 2010); however, food web subnetworks are significantly nested (Kondoh et al., 2010).

These non-random structural patterns are believed to influence ecological dynamics. For example, nestedness may minimise competition and increase biodiversity in mutualistic networks (Bastolla et al., 2009) (but see (James et al., 2012; Staniczenko et al., 2013)), and emerges as a result of an optimisation principle aimed at maximising species abundance in mutualistic networks (Suweis et al., 2013). Modularity is a particularly important property, because it is related to robustness (Hintze and Adami, 2008) and evolvability (Yang, 2001). In addition to species diversity (the number of organisms) and connectance (the relative number of interactions), both modular and nested architectures are also related to ecosystem stability (i.e. persistence and resilience) (Thébault and Fontaine, 2010). However, the relative contributions of nestedness and modularity, in addition to the effects of diversity and connectance, to ecosystem stability differs between mutualistic and antagonistic networks.

Although the structure and stability of ecological networks is highly significant in ecology, the impact of the environment (e.g. geography and climate) on ecological communities is equally significant, because it is important when discussing the effect of climate change on ecosystems. In particular, the environment is expected to influence ecological networks according to latitudinal gradients in species diversity (Araújo and Costa-Pereira, 2013; Condamine et al., 2012). In fact, several studies (e.g. (Baiser et al.,

2012; Marczak et al., 2011)) have focused on latitudinal and geographic variations in ecological networks, although these studies do not mention large-scale structural patterns such as modularity and nestedness.

Because of the importance of modularity and nestedness, the association between these structural patterns and the environment is a key area for active investigation. Trøjelsgaard and Olesen (Trøjelsgaard and Olesen, 2013) found that the mean annual precipitation affects both nestedness and modularity in pollination networks, independent of the sampling effort. Dalsgaard et al. (Dalsgaard et al., 2013) demonstrated that historical climate change (i.e. the quaternary rate of temperature change) was negatively associated with modularity and positively associated with nestedness in pollination networks. However, temperature seasonality, unrelated to historical climate change, exhibits a positive correlation with modularity in seed dispersal networks, using a more realistic definition of modularity (Schleuning et al., 2014), whereas temperature change rate and phylogenetic signals are only weakly associated with modularity.

The positive correlation observed between climatic seasonality and modularity is consistent with several theoretical studies (Friedlander et al., 2013; Kashtan and Alon, 2005), which demonstrate that modular networks spontaneously evolve under changing environments, using an evolutionary optimisation algorithm based on edge rewiring (mutation). Lipson et al. (Lipson et al., 2002) suggested that environmental variability can lead to modularity. Several data analytical studies have found a positive correlation between environmental variability and network modularity in several types of biological system (e.g. metabolic networks (Parter et al., 2007) and cancer signalling networks (Takemoto and Kihara, 2013)). Nevertheless, scepticism still exists regarding the impact of environmental variability on modularity in intracellular networks (Clune et al., 2013; Hansen, 2003; Holme, 2011; Takemoto, 2013, 2012).

Because of the generality of this theory, changes in network structure due to environmental variability (increases in modularity, in particular) such as climatic seasonality should also be investigated in different types of ecological networks. However, the relationship between the environment and ecological network structure is not well understood. This hypothesis has only been partially supported in plant-seed dispersal networks, and environment-structure relationships have only been investigated in plant-animal mutualistic networks (i.e. pollination and seed dispersal networks). Therefore, in this study, we conducted a detailed investigation on the relationship between the environment and ecological network structure by using the data on a number of empirical ecological networks collected from the literature and from databases. In particular, the climatic seasonality effects on (plant-pollinator) mutualistic networks, in addition to environment-structure relationships in food webs and mutualistic networks were evaluated, because they are still poorly understood. In addition, we compared the environment-structure relationships between food webs and mutualistic networks, and discussed the contribution of such relationships to ecosystem stability.

## 2. Materials and Methods

### 2.1 Construction of ecological networks

Food web data were downloaded from the GlobalWeb database (Thompson et al., 2012) (www.globalwebdb.com). Plant-pollinator mutualistic network data were obtained from the Supporting Online Material (Database S1) in (Bascompte et al., 2006), and the Interaction Web DataBase (www.nceas.ucsb.edu/interactionweb/). After removing duplications, we selected ecological networks, the locations (i.e. latitude and longitude) of which could be identified in the literature: 305 food webs and 50 plant-pollinator networks were found.

We constructed ecological networks according to adjacency matrices or lists of species interactions provided by the databases. Food webs are represented as networks, in which nodes and edges correspond to organisms and trophic links (Baiser et al., 2012; Kondoh et al., 2010; Thompson et al., 2012). It is to be noted that food webs are represented as unipartite directed networks, because predator-prey relationships are direction-oriented. However, plant-pollinator networks are represented as bipartite networks, because mutualistic links are only found between two types of organisms (i.e. plants and animals) (Bascompte et al., 2003; Olesen et al., 2007; Trøjelsgaard and Olesen, 2013). Both types of ecological network were represented as binary networks (i.e. presence 1, or absence 0, of a given link), because the databases partially included binary data, and the software programs needed for calculating modularity and nestedness require binary networks.

### 2.2 Climatic parameters and elevation

We manually extracted the information on the latitudes and longitudes of the ecological networks from the literature. Consequently, we obtained the following climate data, with a spatial resolution of 0.5 min of a degree (i.e. $0.93 \times 0.93 = 0.86$ km$^2$), from the WorldClim database (version 1.4, release 3) (Hijmans et al., 2005) (www.worldclim.org) using R version 3.0.2 (www.R-project.org) and an R-package **raster** version 2.2-12 (cran.r-project.org/web/packages/raster): annual mean temperature ($T_{mean}$) ($\times$ 10°C), temperature seasonality (standard deviation) ($T_{var}$), annual precipitation ($P_{ann}$) (mm), and precipitation, or rainfall seasonality (coefficient of variation) ($P_{var}$). The WorldClim database defines temperature seasonality as a standard deviation, because the coefficient of variation is nonsensical at temperatures between -1 and +1. Elevations or altitudes (m) were estimated using the Google Elevation Application Programming Interface (API) (developers.google.com/maps/documentation/elevation/).

### 2.3 Network measures

Based on the approach of a previous study (Thébault and Fontaine, 2010), we focused on four types of network parameters: modularity $M$, nestedness NODF, diversity, and connectance. The modularity of networks is often measured using the Newman-Girvan algorithm (Newman, 2006), which is widely used for investigating network modularity (reviewed in (Fortunato, 2010)). Network modularity ($M$) is generally defined as the

fraction of edges that lie within, rather than between, modules, relative to that expected by chance:

$$M = \sum_{s=1}^{N_M} \left[\frac{l_s}{L} - \left(\frac{d_s}{2L}\right)^2\right],$$

where, $N_M$ is the number of modules, $L$ is the number of links in the network, $l_s$ is the number of links between nodes in module $s$, and $d_s$ is the sum of the degrees of the nodes in module $s$. This definition is applicable to directed networks (i.e. food webs) and bipartite networks (i.e. plant-pollinator networks) (Guimerà et al., 2007).

$M$ ranges from 0 to 1; a network with a higher $M$ indicates a higher modular structure. Therefore, the global maximum $M$ must be found over all possible divisions. The maximum $M$ is defined as the network modularity of ecological networks. Since it is difficult to find the optimal division with the maximum $M$, approximate optimisation techniques are required (Fortunato, 2010). An algorithm based on simulated annealing is optimal for finding the maximum $M$, in order to avoid the resolution limit problem in community (or module) detection as much as possible (Fortunato and Barthélemy, 2007; Fortunato, 2010). Therefore, in the present study, we used BIPARTMOD software (Guimerà et al., 2007) (etseq.urv.cat/seeslab/downloads/bipartite-modularity) for calculating the modularity of directed and bipartite networks, based on simulated annealing.

Nestedness was evaluated using the NODF score (Almeida-Neto et al., 2008), which ranges from 0 to 100. Although several definitions of nestedness have been proposed, the NODF score does not overestimate the degree of nestedness, which occurs with other definitions of nestedness (Almeida-Neto et al., 2008). NODF is related to the proportion of shared interactions between species pairs over a bipartite network (see (Almeida-Neto et al., 2008) for details). In particular, we used the function nestednodf, available in the R package **vegan** version 2.0-9 (CRAN.R-project.org/package=vegan). Since food webs are unipartite networks, we represented them as resource-consumer bipartite networks when calculating NODF, in the same manner as in a previous study (Bascompte et al., 2003).

We also considered the two simplest measures of network complexity: diversity and connectance. These parameters have been studied in terms of structure-stability relationships since the early 1970s (see (Allesina and Tang, 2012; Bascompte, 2010; May, 1972; Mougi and Kondoh, 2012; Thébault and Fontaine, 2010)). Diversity is defined as the number of organisms in a network. The diversities of a food web that consists of $S$ organisms, and plant-pollinator networks with $P$ plants and $A$ pollinators, are defined as $S$ and $P + A$, respectively. Connectance is similar to graph density (i.e. the ratio of the number of edges and the number of possible edges in a network). The connectances of food webs and plant-pollinator networks are defined as $L_t/[S(S - 1)]$ and $L_m/[PA]$, respectively. $L_t$ and $L_m$ correspond to the number of trophic links in a food web and the number of mutualistic links in a plant-pollinator network, respectively.

## 2.4 Networks investigated in this study

For an appropriate evaluation of environment-structure relationships, we investigated a part of the ecological networks (i.e., 305 food webs and 50 plant-pollinator networks), constructed according to the above procedure, for the following reasons. Because of the possibility that the calculation methods underestimate the degree of structural patterns, we only extracted non-random (i.e. significantly modular and/or nested) networks, in the same manner as in a previous study (Trøjelsgaard and Olesen, 2013). Module detection algorithms are known to have a resolution limit problem (Fortunato and Barthélemy, 2007; Fortunato, 2010): the modularity in poorly-sampled or small networks tends to be undervalued; therefore, we used a simulated annealing-based algorithm. Several previous studies (Flores et al., 2011; Olesen et al., 2007; Schleuning et al., 2014) have also identified a similar problem in module detection. In contrast, the NODF score strongly depends on the matrix fill (i.e. connectance) (Almeida-Neto et al., 2008); therefore, nestedness may be underestimated or overestimated. Consequently, we cannot claim that the ecological networks were randomly constructed in terms of modularity and/or nestedness, or whether the significance of structural patterns was affected by technical limitations in the algorithms used for calculating the degree of structural patterns.

A comparison with null model networks was used to evaluate statistical significance. For food webs (i.e. unipartite directed networks), we used networks generated by an edge-switching algorithm (Milo et al., 2002). This algorithm generates a random network by rewiring two randomly selected edges, until the rewiring of all edges is completed and the number of unidirectional and bidirectional edges are preserved (see the supporting online material in (Milo et al., 2002) for details). For plant-pollinator networks (i.e. bipartite networks), the null model II (Bascompte et al., 2003) (essentially similar to the fixed row-fixed column null model, used in (Almeida-Neto et al., 2008)) was used. This null model generates random bipartite networks with degree sequences that are similar to those of real networks. In the null model, the probability that a plant connects to an animal is proportional to the product of node degrees for a given plant or animal.

The significance of network measure $X$ was evaluated using the Z-score: $Z_X = (X_{real} - X_{null})/SD_{null}$, where $X_{real}$ is a network measure (i.e. $M$ or NODF) of a real-world network. $X_{null}$ and $SD_{null}$ are the average network measure value and the standard deviation, respectively, obtained from 500 null model networks constructed using the above algorithms. In this study, we considered the ecological networks with $Z_M > 2$ and/or $Z_{NODF} > 2$ (i.e. $p < 0.05$ using the Z-test). We found that the subsequent results were very similar in the case of $Z_X > 2.6$ (i.e. $p < 0.01$ using the Z-test).

A number of factors affect ecological network structure. Therefore, we considered differences in interaction and ecosystem type. Food webs that include parasites have two types of interaction: trophic interactions and host-parasite interactions. However, network analyses do not explicitly consider such a difference in interaction type. To avoid the difficulty of interpreting such data, we only selected food webs without parasites, according to the classification of the GlobalWeb database.

Food webs were classified as several ecosystem types: freshwater, marine, terrestrial, and estuarine, according to the classification of the GlobalWeb database. We found 94 freshwater, 31 marine, 17 terrestrial, and 12 estuarine food webs by using the above selection criteria. We chose freshwater food webs for data analysis because of the number of samples. Although there may have been a sufficient number of marine food webs for analysis, we omitted them because the WorldClim database (Hijmans et al., 2005) mainly focuses on global land area.

To reduce data redundancy as much as possible, duplications of observation sites were eliminated. In particular, the largest ecological networks (i.e. those with the largest number of organisms) were selected when there were multiple ecological networks with similar observation sites. We, therefore, included 67 freshwater food webs and 29 plant-pollinator networks.

2.5 Statistical analysis

For measuring statistical dependence between the geographical parameters and network measures, we used the Spearman's rank correlation coefficient $r_s$, which is a non-parametric measure (i.e. it is relatively robust to outliers and can also evaluate nonlinear relationships).

To evaluate the contribution of each geographical parameter to a network measure, we conducted kernel partial least squares (PLS) regression analysis (Rosipal and Trejo, 2001), using R version 3.0.2 with its function plsr, which is available in the R package *pls* version 2.3-0 (Mevik and Wehrens, 2007). PLS regression analyses are more powerful than other multivariate analysis techniques because they can identify significant effects on the dependant variables (i.e. $M$, NODF, diversity, and connectance) from a large number of explanatory variables (i.e. latitude, annual mean temperature, temperature seasonality, annual precipitation, rainfall seasonality, and elevation) when a high level of multicollinearity is observed among the explanatory variables.

In particular, $r_s$ (or coefficient of determination: $r_s^2$) and the proportion of variance explained in PLS indicate the validity of statistical models.

## 3. Results

3.1 Distribution of ecological networks

The ecological networks included in this study were obtained from across the world (Figure 1A), and were uniformly sampled from several countries (Figures 1B and 1C). These networks were therefore suitable for evaluating environment-structure relationships with climate. The ecological networks selected may be slightly biased, due to the researchers' interests. For example, the proportion of food webs from New Zealand is larger than that from other countries. This may be because the database on food webs (i.e. GlobalWeb) is hosted by the University of Canberra, Australia. In contrast, the pollination networks are weighted towards Japan. This may be because Japanese research groups have published large-scale studies on pollination networks for many years. Note that these biases do not indicate data redundancy; such redundancy was reduced as much as possible by the data selection (see Section 2d).

The number of pollination networks in this study is lower than that in a previous study (Trøjelsgaard and Olesen, 2013), because the previous study included unpublished data. However, this difference poses few problems, because the statistical tendency is similar between this study and the previous study.

## 3.2 Rainfall seasonality affects freshwater food web structure

The correlation analysis found that rainfall seasonality was associated with network measures (Table 1A), except for NODF. Although some outliers are observed, this observed relationship is probably reliable because of the use of suitable statistical methods (see Section 2.5). Rainfall seasonality (i.e. variability) increased connectance and decreased diversity. Remarkably, the network modularity $M$ of freshwater food webs was positively correlated with rainfall seasonality (Figure 2A), which is expected, if increases in modularity are due to environmental variability (Friedlander et al., 2013; Kashtan and Alon, 2005). However, it was only weakly correlated with temperature seasonality ($r_s = -0.23$; $p = 0.26$).

Since connectance mainly affects $M$ from a theoretical perspective (Guimerà et al., 2004), it is possible that the positive correlation observed between $M$ and rainfall seasonality was caused by a relationship between connectance and rainfall seasonality. However, such an effect may not occur for the following reasons. The theory predicts that $M$ decreases with connectance; therefore, a negative correlation should be found between $M$ and rainfall seasonality. However, the opposite was observed. In addition, there was no correlation between $M$ and connectance ($r_s = 0.17$; $p = 0.15$). To avoid the effects of connectance on $M$, a normalised modularity is often used (Holme, 2011; Schleuning et al., 2014), and is defined as $M_n = (M_{real} - M_{null})$. $M_n$ also exhibited a positive correlation with rainfall seasonality ($r_s = 0.31$; $p = 0.0098$). These results suggest that the positive correlation observed between $M$ and rainfall seasonality is independent of the other network parameters.

Although a negative correlation between NODF and latitude was found (Table 1A), there is a possibility that this was an indirect effect, because NODF is known to increase with increasing connectance (Almeida-Neto et al., 2008). Indeed, a positive correlation between NODF and connectance was observed ($r_s = 0.34$; $p = 0.0043$). To avoid any bias due to connectance, a normalised score of NODF based on the $Z$-score (i.e. $Z_{NODF}$) is often used (Almeida-Neto et al., 2008). A correlation between $Z_{NODF}$ and latitude was not found ($r_s = -0.059$; $p = 0.64$). We therefore can conclude that no geographical parameters contributed to nestedness.

The importance of rainfall seasonality on food web structure is still debated, because of the correlation between geographical parameters. For example, there was a strong positive correlation between rainfall seasonality and annual mean temperature ($r_s = 0.90$;

$p < 2.2 \times 10^{-16}$). To reduce such multicollinearity among the geographic parameters as much as possible, we conducted a PLS regression analysis. The loadings plot of the objective variable $M$ (Figure 2B) indicates that rainfall seasonality had the largest loading (in absolute number) of the first principle component, suggesting that rainfall seasonality was the main contributor to network modularity. However, the annual mean temperature had the third largest loading (in absolute number), and there were no major differences between the loadings (in absolute number) of rainfall seasonality (0.60) and annual mean temperature (0.54); therefore, it indicates that the annual mean temperature (secondly) contributed to food web structure. Diversity and connectance exhibited a similar pattern. In both cases, the variable with the largest loading (absolute value) of the first principle component was rainfall seasonality (diversity: -0.66, 41.4% variance explained; connectance: 0.64, 26.6% variance explained); the other geographic parameters had lower loadings.

### 3.3 Temperature seasonality influences mutualistic network structure

As with food webs, simple statistical analysis showed that plant-pollinator mutualistic network structure was influenced by temperature seasonality more than rainfall seasonality (Table 1B). In particular, increases in temperature seasonality resulted in increases in diversity and nestedness, and decreases in connectance.

Although the correlation between modularity $M$ and temperature was not conclusive ($r_s$ = 0.35; $p$ = 0.069), this positive relationship is consistent with the hypothesis that the increases in modularity are due to environmental variability. However, it remains possible that this correlation is an artefact, because a negative correlation between $M$ and connectance, as predicted by theory (Guimerà et al., 2004), was observed ($r_s$ = -0.72; $p$ = $3.0 \times 10^{-5}$), and a correlation between the normalised modularity $M_n$ and temperature seasonality was not observed ($r_s$ = 0.14; $p$ = 0.48). A similar pattern was observed in seed dispersal networks, which are a different type of mutualistic network. Climatic seasonality has little effect therefore on network modularity; however, this is still debatable because it may depend on the definition of modularity (see Section 4). As reported in a previous study (Trøjelsgaard and Olesen, 2013), annual precipitation exhibited a positive correlation with $M$ (Table 1B) and $M_n$ ($r_s$ = 0.56; $p$ = 0.0018). In addition, a negative correlation between NODF and annual precipitation was observed in the same study, but it was not statistically significant ($r_s$ = -0.34; $p$ = 0.073). However, the results of the present study do not contradict previous results, because the effect size (i.e. squared correlation coefficient) was similar between this study (0.12) and the previous study (0.10). The lower $p$ value in this study may be because the number of samples included (28) was lower than that in the previous study (48). Although a negative correlation between NODF and temperature seasonality was found (Table 1B), it remains possible that the observed association was an artefact caused by the positive correlation between NODF and connectance (Almeida-Neto et al., 2008). As expected, a strong positive correlation between NODF and connectance was observed ($r_s$ = 0.91; $p$ = $5.8 \times 10^{-7}$), which may have caused the observed negative correlation between NODF and temperature seasonality, because connectance was also negatively correlated with temperature seasonality (Figure 3A). In addition, there was no significant relationship between the normalised score of NODF (i.e. $Z_{NODF}$) and temperature seasonality ($r_s = -0.20$; $p = 0.30$).

Taken together, these results highlight the importance of the effects of temperature seasonality on diversity (Figure 3A) and connectance (Table 1B). The results of the PLS regression analysis suggest that temperature seasonality is the main contributor to diversity (Figure 3B). Changes in diversity depend on animal diversity, because temperature seasonality is correlated with the number of animal species present ($r_s = 0.53$; $p = 0.0036$); however, it is only weakly associated with the number of plant species present ($r_s = 0.36$; $p = 0.060$). Connectance had a similar loadings plot. Temperature seasonality had the largest loading (in absolute number), which was 0.60 for the first principle component (41.0% variance explained).

## *4.* Discussion

This study evaluated the impact of climatic seasonality on freshwater food webs and plant-pollinator networks, inspired by theoretical studies (Friedlander et al., 2013; Kashtan and Alon, 2005), and demonstrated that changes in network structure are caused by environmental variability. As expected from a theoretical study, such an effect of climatic seasonality on network structure was observed in both types of ecological networks.

Interestingly, seasonality parameters that affected network structure differed between food webs and mutualistic networks. In particular, rainfall and temperature seasonality influenced food web structure and mutualistic network structure, respectively. This difference may have been caused by differences in ecosystem type rather than differences in interaction type (i.e. antagonistic or mutualistic). In this study, we focused on freshwater food webs for several reasons (see Section 2.4). These food webs are found in lakes, ponds, streams, reservoirs, etc., and the quantity of water, such as rain precipitation, is expected to influence environmental conditions in these ecosystems. In contrast, pollination systems are generally terrestrial; therefore, they should be mainly affected by temperature rather than precipitation. Of course, other geographical parameters may also influence ecological network structure. For example, temperature may be related to species diversity in ecosystems (Araújo and Costa-Pereira, 2013), and rainfall is also important in pollination systems (González et al., 2009). In fact, temperature seasonality and annual mean temperature are also associated with the diversity and connectance of freshwater food webs, and annual precipitation is related to the modularity of mutualistic networks (Table 1).

A positive correlation between network modularity and rainfall seasonality in freshwater food webs is consistent with the hypothesis that increases in modularity are due to environmental variability (Friedlander et al., 2013; Kashtan and Alon, 2005). Since modularity is related to robustness (Hintze and Adami, 2008), this increase in modularity may be a strategy to increase ecosystem stability under changing environments. A recent theoretical study (Thébault and Fontaine, 2010) has reported that modularity increases resilience (i.e. the speed at which the community returns to equilibrium after a perturbation) in trophic networks, and decreases in diversity and connectance enhance stability. Food webs that experience a high seasonal variation in precipitation have a lower diversity (network size) and a higher modularity, while network complexity (number of links per species: $L_t/S$) is constant, or moderately decreases with rainfall seasonality ($r_s = -0.23$; $p = 0.056$). This pattern is in agreement with theoretical predictions regarding system resilience against environmental perturbation.

Modularity was not significantly correlated with either temperature seasonality or rainfall seasonality in pollination networks, although such a correlation has been found in seed dispersal networks (Schleuning et al., 2014), which is a different type of mutualistic network. This may be because of the difference in the definition of network modularity between this study and the previous study. A recently published study (Schleuning et al., 2014) investigated the modularity of weighted networks (weighted modularity) and found that temperature seasonality was positively correlated with weighted modularity but not with binary modularity, as was used in the present study. Therefore, there may be an association between climatic seasonality and modularity, when considering weighted modularity. Ideally, ecological networks should be represented as weighted networks, because of interaction strengths (e.g. number of visits). However, in this study, pollination networks are represented as unweighted bipartite networks, because the data on pollination networks include binary data. In addition, the definition of weight or interaction strength differs among the data. A more careful examination of the correlation between modularity and climatic seasonality in pollination networks is required althogh the observed significant positive correlation between modularity and annual precipitation at least suggests an effect of climate on pollination network modularity.

The impact of climate on nestedness in pollination networks is debatable. Temperature seasonality exhibited a negative correlation with NODF; however, it was weakly related to the normalised NODF score (Section 3). In addition, mutualistic networks are significantly nested in binary network representations, but not in quantitative (i.e. weighted) network representations (Staniczenko et al., 2013), suggesting that nestedness is an artefact of mutualistic networks. From these results, we can conclude that temperature seasonality mainly affects diversity and connectance in mutualistic networks. In particular, the diversity of pollination networks increased with temperature seasonality (Table 1B), while the number of links per plant or animal species was generally independent of temperature seasonality (per plant, $r_s = -0.25$, $p = 0.18$; per animal, $r_s = -0.26$, $p = 0.17$). This is consistent with theoretical predictions (Thébault and Fontaine, 2010) that higher diversity increases resilience against perturbation and persistence (i.e. the proportion of persisting species once equilibrium is reached).

Nevertheless, note that these results are still debatable because the observed correlations are statistically significant, but weak (i.e., a number of outliers were observed (Figures 2 and 3)). In this case, a possibility of type I error remains. In addition to this, it remains possible that these results were influenced by the following limitations. That is, this study just suggested a possibility of the effect of climatic seasonality on ecological network structure; thus, a careful examination will be required in the future.

The definition of modularity and modules is controversial. The conclusions of this study are limited to unweighted network modularity. However, most functional modules may be defined through module detection methods based on network topology (i.e. in the context of network modularity). It has been reported that the definition of modularity used in this study and many previous studies might not be topologically correct due to locality and limited resolution (Fortunato and Barthélemy, 2007; Fortunato, 2010). Module detection methods for weighted networks (Schleuning et al., 2014) and overlapping communities (e.g. (Ahn et al., 2010; Becker et al., 2012)) may be able to avoid these limitations, because they are more accurate in their prediction of functional modules.

The impact of climatic seasonality on species diversity requires a more careful examination. According to the species-area relationship theory (McGuinness, 1984), diversity depends on the observed area of the ecological network (i.e. diversity increases with increasing area). In this study, we could not extract the relevant information on sampling effort because much of the literature does not clearly describe it. However, we are of the opinion that the effect of sampling effort did not influence the observed results, because there was little variation in sampling effort. Although a positive correlation between observed area and species diversity has been reported in pollination networks, it is not conclusive. The reported effect size (squared correlation coefficient) of the correlation between animal diversity and observed area is 0.09. A previous study (Trøjelsgaard and Olesen, 2013), in which the number $n$ of samples was 49, reported a statistically significant correlation ($p = 0.037$). However, the present study ($n = 28$) did not find such statistical significance ($p = 0.12$). Moreover, this study (Trøjelsgaard and Olesen, 2013) concluded that ecological network structures are independent of sampling effort.

Although the effects of phylogenetic signals should be investigated, we did not consider them in this study, because ecological networks partly consist of species whose descriptions are unknown (i.e. they are expressed as species A and species B) or ambiguous. However, several studies have reported that phylogenetic signals are weak in ecological network structures (Rezende et al., 2007; Schleuning et al., 2014); therefore, we expect that not taking the phylogenetic signals into consideration should not pose any problems. A previous study (Schleuning et al., 2014) also demonstrated that phylogenetic signals are only weakly associated with network structure.

In the context of environmental variability, historical climate change (or rate of climate change) is also important (Loarie et al., 2009). However, we could not evaluate the effect of climate change rate on network structure because the row data of climate change rates were not freely available. However, we expect that ecological network structure is influenced by climatic seasonality rather than by historical climate change, because variability in historical climate change is expected to be lower than that in climatic seasonality. Indeed, a previous study (Schleuning et al., 2014) reported that the rate of climate change did not significantly affect network modularity in seed-dispersal networks. Although a previous study (Dalsgaard et al., 2013) has found an association between modularity and historical climate change (quaternary temperature change rate), it did not consider normalised modularity (i.e. a comparison with null models or random control). Therefore, it remains possible that such an association is an artefact.

In this study, we omitted food webs that included parasites to avoid the difficulty of interpreting the results, because the network measures cannot evaluate differences in interaction type (i.e. trophic links and host-parasite interactions), which are observed in such food webs. However, this does not indicate that parasites are unimportant. On the contrary, parasites play critically important roles in food webs (Dunne et al., 2013); in particular, they increase food web diversity and complexity. In addition, a mixture of interaction types is essential for more realistic interactions, and may contribute to increasing ecosystem stability (Allesina and Tang, 2012; Mougi and Kondoh, 2012). Therefore, network measures for multiplex networks should be considered in the future, to evaluate ecological network structure under more realistic conditions. Multiplex networks consist of a set of nodes connected by different types of link, and have been investigated in the context of social network analysis (Hoang and Antoncic, 2003).

In addition to the above limitations, our analysis has a general limitation, as do many other works on ecological network analyses: there is only a limited knowledge of interspecific reactions (i.e. missing links). To avoid these limitations, more highly normalised databases should be constructed, in addition to finding alternative approaches to network analysis. In particular, information on ecological networks (e.g. ecosystem type, observed site, observed area and time, etc.) should be summarised using an ontological characterisation, such as the Environmental Ontology (EnvO) database (Hirschman et al., 2008), for concise and controlled descriptions of environments.

Although the conclusions are still debatable because of the above limitations, they may enhance our understanding of the effect of climatic seasonality on ecological network structure.

## Acknowledgements

The authors would like to thank Prof. Jennifer Dunne and Prof. Ross Thompson for arranging the use of the GlobalWeb database. This study was supported by a Grant-in-Aid for Young Scientists (A) from the Japan Society for the Promotion of Science (no. 25700030). KT was partly supported by Chinese Academy of Sciences Fellowships for Young International Scientists (no. 2012Y1SB0014) and the International Young Scientists Program of the National Natural Science Foundation of China (no. 11250110508).

# Tables

**Table 1.** Correlations between network measures and geographical parameters in freshwater food webs (A) and plant–pollinator networks (B). $n$ indicates the number of networks used in the analysis. The Spearman's rank correlation coefficient $r_s$ and the associated $p$ values are presented. As representative examples, only statistically significant correlations ($p < 0.05$) are included.

|  | Network measure | Geographical parameter | $r_s$ | $p$ |
|---|---|---|---|---|
| (A) $n = 67$ | Modularity | Annual mean temperature | 0.37 | 0.0020 |
| | | Rainfall seasonality | 0.41 | 0.00048 |
| | NODF | Latitude | 0.30 | 0.015 |
| | Diversity | Annual mean temperature | -0.52 | $4.8 \times 10^{-6}$ |
| | | Temperature seasonality | -0.37 | 0.0022 |
| | | Annual precipitation | 0.44 | 0.00020 |
| | | Rainfall seasonality | -0.61 | $5.8 \times 10^{-8}$ |
| | Connectance | Annual mean temperature | 0.32 | 0.0079 |
| | | Temperature seasonality | 0.29 | 0.016 |
| | | Annual precipitation | -0.38 | 0.0014 |
| | | Rainfall seasonality | 0.40 | 0.00072 |
| (B) $n = 28$ | Modularity | Annual Precipitation | 0.50 | 0.0068 |
| | NODF | Temperature seasonality | -0.45 | 0.017 |
| | Diversity | Temperature seasonality | 0.51 | 0.0047 |
| | Connectance | Temperature seasonality | -0.48 | 0.010 |

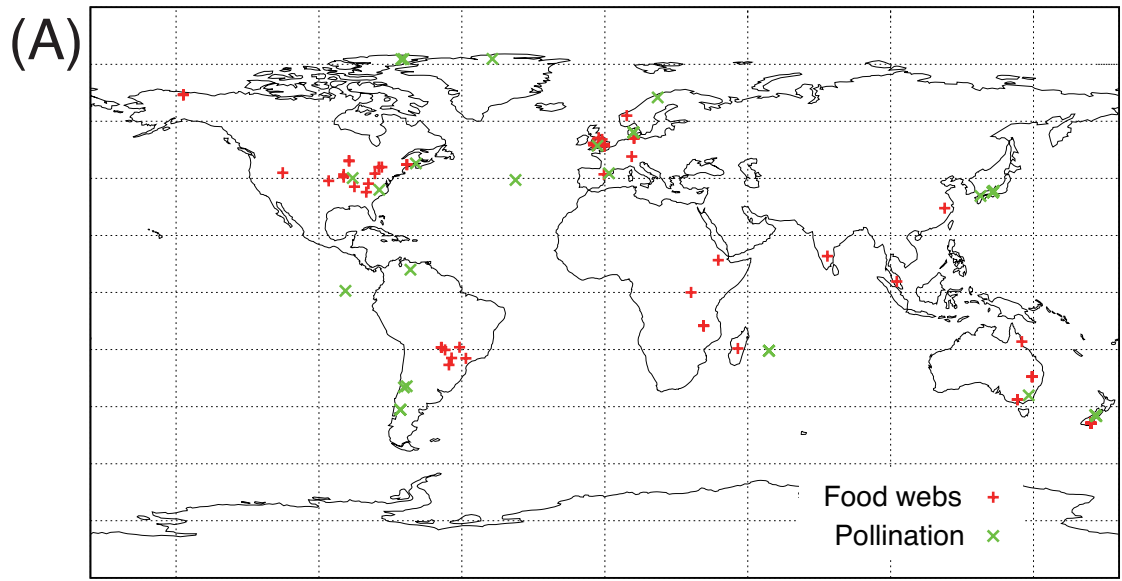
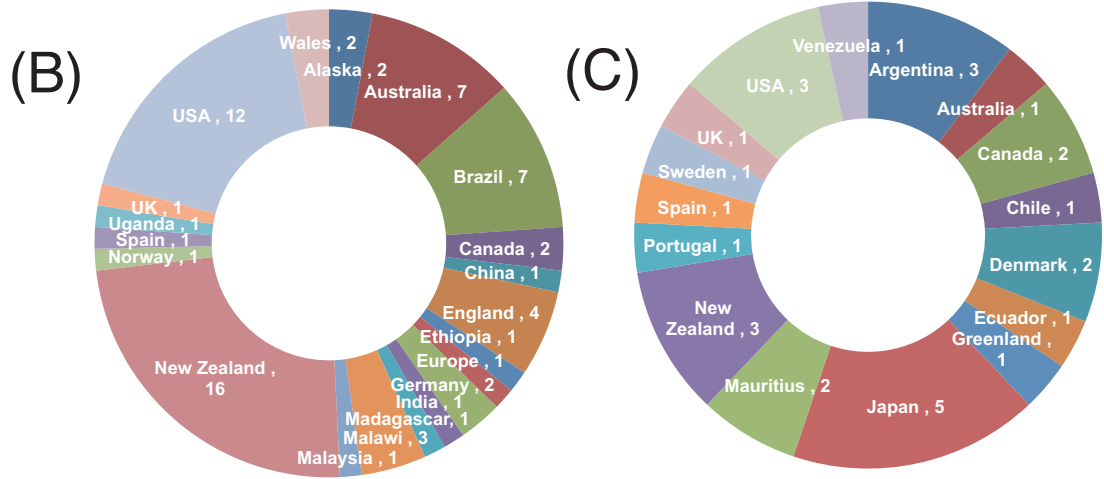

**Figure 1.** Distribution of the ecological networks used in this study on the world map (A), statistics of the observed ecological networks, according to country, in food webs (B), and plant-pollinator mutualistic networks (C). The numbers in the circle Figures correspond to the number of observed networks in that country.

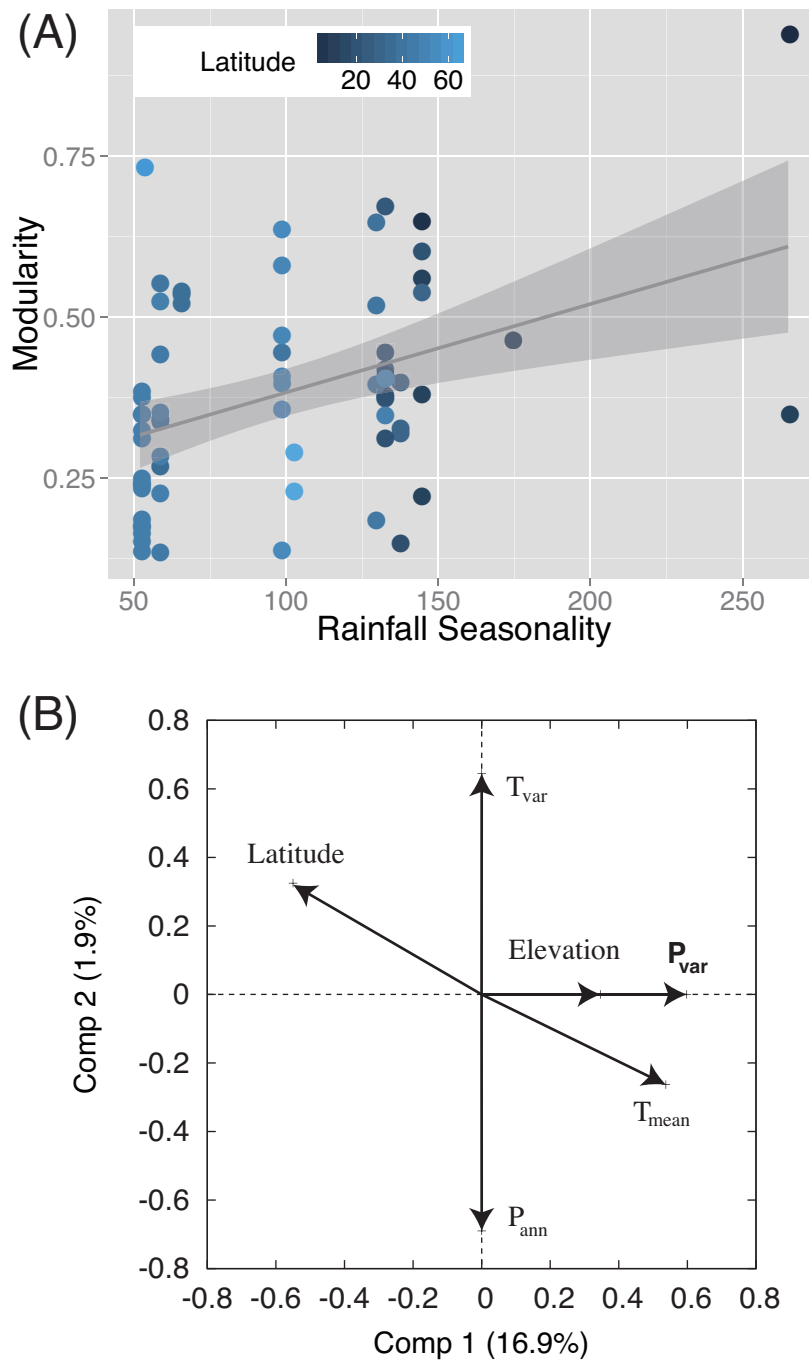

**Figure 2.** (A) The observed positive correlation between network modularity and rainfall seasonality in freshwater food webs (Spearman's rank correlation coefficient $r_s = 0.41$; $p = 0.00048$). The regression line (grey solid line) is based on a simple linear regression analysis. The grey shading indicates the confidence interval around the mean standard error. (B) Loadings plot from the partial least square regression analysis. The percentages in parentheses correspond to the proportion of variance explained by components 1 and 2. The geographical parameter indicated in boldface had the largest loading (in absolute number) of the first component (Comp 1).

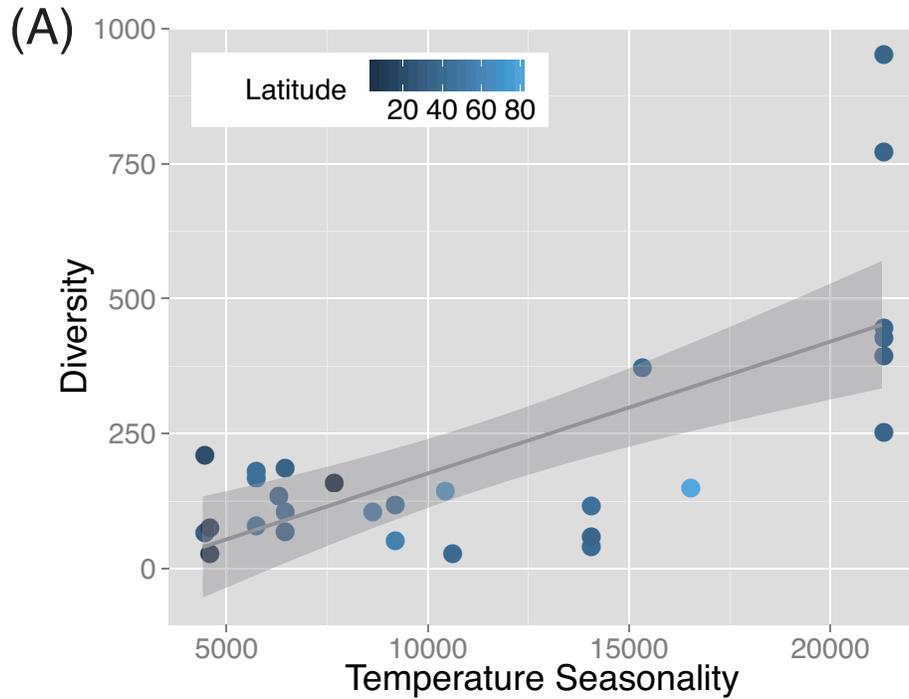

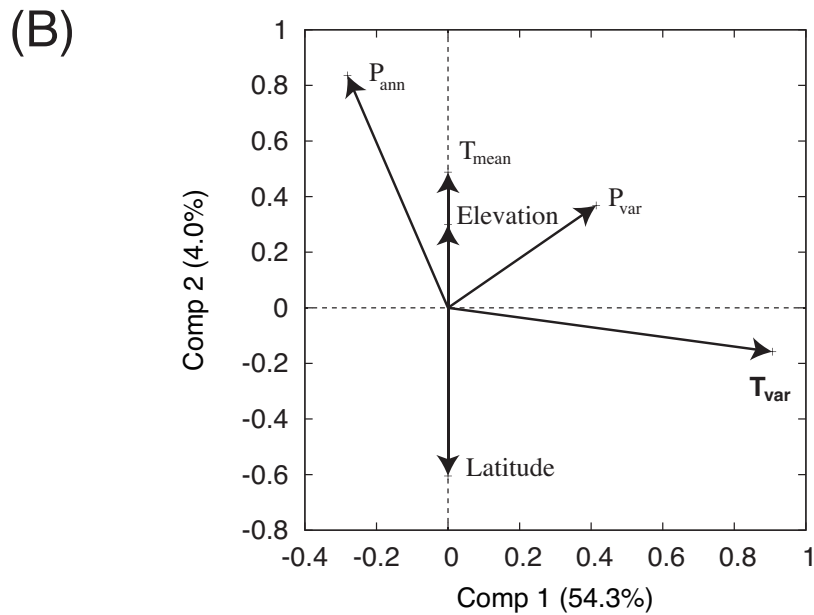

**Figure 3.** (A) The observed positive correlation between diversity and temperature seasonality in plant-pollinator mutualistic networks (Spearman's rank correlation coefficient $r_s = 0.51$; $p = 0.0047$). The regression line (grey solid line) is based on a simple linear regression analysis. The gray shading indicates the confidence interval around the mean standard error. (B) Loadings plot from the partial least square regression analysis. The percentages in parentheses correspond to the proportion of variance explained by components 1 and 2. The geographical parameter indicated in boldface had the largest loading (in absolute number) of the first component (Comp 1).